\documentclass[prd,nofootinbib,twocolumn,superscriptaddress]{revtex4}
\usepackage{graphicx}
\usepackage{bm}
\usepackage{subfigure}

\begin{document}
\title{Degree-scale anomalies in the CMB: localizing the first 
peak dip to a small patch of the north ecliptic sky}

\author{Amanda Yoho}
\affiliation{
CERCA, Department of Physics, Case Western Reserve University,
10900 Euclid Avenue, Cleveland, OH 44106-7079, USA.}
\author{Francesc Ferrer}
\affiliation{
Physics Department and McDonnell Center for the Space Sciences, 
Washington University, St Louis, MO 63130, USA}
\author{Glenn D. Starkman}
\affiliation{
CERCA/ISO, Department of Physics, Case Western Reserve University,
10900 Euclid Avenue, Cleveland, OH 44106-7079, USA.}
\begin{abstract}
\noindent Noticeable deviations from the prediction of the fiducial LCDM
 cosmology are found in the angular power spectrum of the CMB. Besides
 large-angle anomalies, the WMAP $1^{st}$ year data revealed a dip in the
 power spectrum at $l \sim 200$, which seemed to disappear in the $3^{rd}$ year
 and subsequent angular power spectra. Using the WMAP single $1^{st}$, $3^{rd}$, and $5^{th}$ year 
 data as well as the total $5$ year coadded data, we study the intensity and
 spatial distribution of this feature in order to unveil its origin and
 its implications for the cosmological parameters.  We show that in all 
  WMAP data releases there is a substantial suppression of the first Doppler peak
  in a region  near the north ecliptic pole.
\end{abstract}

\maketitle

\section{Introduction}
The dynamics of both individual galaxies and clusters of galaxies, the Hubble
diagram for distant Type Ia supernovae and the pattern of temperature 
anisotropies in the Cosmic Microwave Background (CMB) broadly support the standard
flat $\Lambda$-dominated cosmological model with a nearly scale-invariant
spectrum of adiabatic Gaussian primordial fluctuations, 
such as might be generated at the end of an inflationary epoch, 
as an accurate description of our universe. 

In particular, this simple model provides an acceptable fit to the 
Wilkinson Microwave Anisotropy Probe (WMAP) observations of the CMB temperature
anisotropies. The pattern of acoustic peaks and troughs 
predicted by the model to be imprinted on the angular power spectrum by the
primordial fluctuations  in the inflationary field and its subsequent
evolution has been detected and characterized to good precision~\cite{
Hinshaw:2003ex,Spergel:2003cb,Hinshaw:2006ia}.
A global fit to the data has enabled cosmologists to determine the handful of parameters
of the inflationary $\Lambda$CDM model, which does a remarkably good job at describing the
spectrum.

The WMAP data, however, also exhibits
unexpected anomalies that have sparked considerable attention \cite{Bennett:2010jb,Copi:2010na}.
These include excesses or deficiencies of power in at least three bins --
 $\ell \sim 22$, $44$ and $200$ --  violations of statistical isotropy
at $\ell<6$~\cite{de OliveiraCosta:2003pu,Schwarz:2004gk,Land:2005ad}, and 
a severe  lack of large angle  correlations requiring the low-$\ell$ $C_\ell$
to not be independent~\cite{Spergel:2003cb,Copi:2006tu}.
Hemispheric asymmetries in the temperature power spectrum extending over a wide range of angular scales 
as well as hemispheric dependent non-gaussian signatures were
reported in~\cite{Eriksen:2003db,Hansen:2008ym,Rath:2007ti,Rossmanith:2009cy}, and a non-Gaussian cold spot in the
southern galactic hemisphere was found using wavelet 
transforms~\cite{Vielva:2003et,Cruz:2004ce}. Evidence for primordial non-Gaussianity
of the local type in the temperature spectrum in the
third year data has been 
claimed in~\cite{Yadav:2007yy} (However, it does not seem to be
present when later data are included in the 
analysis~\cite{Komatsu:2008hk,Smith:2009jr}).

If these anomalies are of primordial origin, rather than caused by 
unaccounted foreground or secondary effects, there could be profound
implications for our understanding of the early Universe. 
For example, while the simplest single field inflationary scenarios predict negligible deviations from
Gaussianity, larger deviations can be accommodated in extended particle 
physics models~\cite{Bartolo:2004if}.
The lack of power in some low-$\ell$ multipoles, 
though not the multipole alignments and not the
lack of large-angle correlations, could be 
explained by the presence of anticorrelated isocurvature perturbations
such as those expected in the curvaton model~\cite{Gordon:2003fi,Ferrer:2004nv}.
Topological defects could be at the origin of the  cold spot~\cite{Cruz:2007pe}.

Most of these anomalies were first noted in the first year WMAP data
release and persisted into the subsequent data 
releases~\cite{Copi:2006tu,Eriksen:2007pc}. 
The dip in the first peak found in the first year WMAP data release is believed
to be a notable exception -- it is no longer present when the newly adopted 
noise weighting scheme  is used to extract the best fit power spectrum. 
This anomaly in the first year angular power spectrum was, 
thus, attributed to a noise fluctuation~\cite{Hinshaw:2006ia}. 

Surprisingly, the dip in the power spectrum around $\ell \sim 200$ disappears when
data from the {\em ecliptic} poles is not included in the analysis of the 1st
year WMAP data regardless of the weighting scheme
(see Fig. 7 in~\cite{Hinshaw:2003ex}). 
Archeops,  the sole previous ground-based experiment that observed the region of the sky
around the  north ecliptic pole, showed a dip around the first peak. 
All other similar experiments showed no such dip and had access only
to southern ecliptic regions~\cite{Wang:2002rta}.
This suggests that the differences might not simply be due to a 
noise fluctuation, but might rather reflect a real anomaly.

In this paper we take a closer look at the degree-scale CMB temperature anisotropies. 
We estimate the primordial power spectrum  using data from different regions of the sky 
according to the two weighting schemes that have been used by the WMAP collaboration. 
Our findings show that there is a substantial suppression of the first Doppler peak
in a region near the north ecliptic pole. This reduction in the temperature
power spectrum around the first acoustic peak occurs regardless of
the weighting scheme used, although the detailed shape (i.e. the dip like
feature) is only present when the analysis follows the strategy used
by WMAP in their first year release and might also depend on the particular
binning used to present the data.

The rest of the paper is organized as follows. In section~\ref{sec:method},
we review the method used for estimating the power spectrum, $C_{\ell}$, from
observations.
%, emphasizing the different strategies that have been used to
%maximize the signal-to-noise ratio in an unbiased way. 
Section~\ref{sec:data} describes the datasets that we used, together
with the processing steps taken to minimize the influence of non-cosmological
foregrounds. In section~\ref{sec:stat} we discuss the 
method used to determine the statistical significance of our analysis of 
the data. Next, in section~\ref{sec:results}, before our conclusions, we
present the results of our analysis and briefly comment on implications
for the physics underlying the CMB. 

\section{Estimation of power spectra}
\label{sec:method}

Two main approaches have been used to estimate the power spectrum from a CMB
data map, including sources of errors and partial sky coverage, namely
maximum-likelihood estimators and pseudo-$C_{\ell}$ (PCL) methods. The WMAP team
used the latter strategy for the analysis of their first year data, but
changed to a hybrid estimator when producing the third year results,
in agreement with the discussion in~\cite{Efstathiou:2003dj}. 
The differences, however, concern only the lowest multipoles ($\ell \leq 30$), 
which are not the object of our study. Hence, we base our results
on the pseudo-$C_{\ell}$ technique~\cite{Hivon:2001jp}, which we now briefly
review.

The usual decomposition of a sky map $\Delta T (\mathbf{n})$ in spherical
harmonics can be generalised to the case of partial sky coverage by introducing
a position-dependent weight, $W(\mathbf{n})$, which is set to zero in the
regions of the sky that are to be excluded from the analysis:
\begin{equation}
\tilde{a}_{\ell m}^i = \Omega_p \sum_p {\Delta T^i (p) W^i (p) Y^*_{\ell m}(p)}\, .
\label{eq:walm}
\end{equation}
Here the map has been discretized in pixels 
sub-tending a solid angle $\Omega_p$, and the index $i$ refers 
to the particular differential assembly (DA) under scrutiny (see~\cite{Hinshaw:2003ex} 
for details on the datasets).

Even though the pseudo-power spectrum, $\tilde{C}_{\ell}$, derived
from weighted maps~(\ref{eq:walm}), 
\begin{equation}
\tilde{C}_{\ell} = \frac{1}{2 \ell +1} \sum_{m=-\ell}^{\ell}{\left| \tilde{a}_{\ell m}\right|^2},
\label{eq:pscl1}
\end{equation}
differs from the full-sky angular spectrum, $C_{\ell}^{\rm sky}$,
their ensemble averages are related by a mode coupling matrix,
$G_{\ell \ell '}$, which can be inverted to obtain an 
estimator of the power spectrum:
\begin{equation}
\left< C_{\ell}^{\rm sky} \right> = C_{\ell} = \sum_{\ell'} {G_{\ell \ell '}^{-1} \tilde{C}_{\ell '}}.
\label{eq:qest}
\end{equation}
The coupling matrix depends only on the 
geometry of the weight function, $W(\mathbf{n})$.
An expression for $W(\mathbf{n})$ can be found in~\cite{Hivon:2001jp}.

In a multichannel experiment like WMAP, the noise between two different DAs is
assumed to be uncorrelated. With a straightforward generalization of 
Eq.~(\ref{eq:pscl1}) for the analysis of two DA maps at a time,
\begin{equation}
\tilde{C}_{\ell} = \frac{1}{2 \ell +1} \sum_{m=-\ell}^{\ell}{\tilde{a}_{\ell m}^i 
\tilde{a}_{\ell m}^{j*} },
\label{eq:pscl}
\end{equation}
we obtain an estimate of the true power spectrum that is not biased
by noise. 

The weight function, $W(\mathbf{n})$, is zero
where the foreground emission cannot be reliably eliminated. The main sources
of non-cosmological origin (free-free, dust and synchrotron emission) are
subtracted from each DA map by fitting templates and taking advantage of the
different frequency dependence of each signal. The intensity of the
foregrounds in the plane of the Galaxy, however, disallows a proper separation
of the cosmological component in this region, which should be excluded
from our analysis. We will set $W$ to zero in the region specified by the 
KQ85 five year mask, an area of roughly 15\% of the sky.

It is also advantageous to choose a different form of $W(\mathbf{n})$
in those regions where it is non-zero,
depending on the multipole angular scale involved. Maximum-likelihood
estimators, leading to smaller error bars, weight the data by the
inverse covariance matrix, $C^{-1}=(S+N)^{-1}$, where
$S$ and $N$ are the contributions from the signal and the instrument noise respectively. 
We can mimic this optimal
sampling for the PCL estimator by using a unit uniform weight in the
signal dominated, $C^{-1} \approx S^{-1}$, low $\ell$ regime. An inverse
noise weight, $W^p \approx N_{obs}^p$, is more suitable for the
noise dominated high multipoles~\cite{Efstathiou:2003dj}. 
In the first-year WMAP analysis, the two regimes were taken to be $\ell <200$ and $\ell >450$, 
while a transitional weighting interpolation was used for the intermediate multipoles~\cite{Verde:2003ey}. 
For the third year analysis, the WMAP team used a sharp
transition between the uniform weight and 
inverse noise weight at $\ell =500$ (the transition was made
at $\ell=600$ for the seventh year analysis, although this does not
affect our discussion). 
The fact that the dip-like feature in
the first acoustic peak is not evident in the WMAP third year plot, lead
the WMAP collaboration to conclude that it was largely due to the sharp
weight transition at $\ell <200$. We will see in the following that, in
addition to this effect, the 
CMB power spectrum at these scales seems to vary markedly among
different regions of the sky.

\section{Data sets}
\label{sec:data}

The WMAP team obtained its best estimate power spectrum by optimally weighting
all the possible pairings of Q, V and W DAs, although the Q band was dropped
in the third year estimate, since it is the most prone to foreground and diffuse
source contamination.
%\footnotemark[1] 
Our aim here is not to obtain the best power spectrum
estimate, but to study how it changes when looking at different parts of
the sky or when using different weighting schemes. We thus obtain our
power spectrum estimate by applying Eq.~(\ref{eq:qest}) to the
foreground reduced V and W frequency band maps for the single first, third, and fifth 
year maps as well as the complete fifth year coadded map. We perform our analysis using the data processing method 
corresponding to each data release year as well as the applying the first and third year
weighting schemes~\cite{Hinshaw:2003ex,Hinshaw:2006ia} to all datasets.

Oddly enough, some of the anomalies found at large angular
scales correlate with the ecliptic~\cite{Schwarz:2004gk}, and the
WMAP team first year data gives a different angular power spectrum around $\ell \sim 200$ 
in the ecliptic plane than at the poles~\cite{Hinshaw:2003ex}.
A random search in all directions could potentially
reveal a more significant region, however when
such an analysis was performed for the 1st year data the results also
pointed to the ecliptic poles, as Eriksen, et al. found significantly low power in the northern
hemisphere near the ecliptic pole for the range $\ell=2 - 40$~\cite{Eriksen:2003db}.
We, thus, choose to examine two {\it a priori} selected 30 degree diameter spherical caps 
centered on the ecliptic poles. A later analysis (also on the first year data) that computed the directional 
dependence of the temperature
power spectrum for an $\ell$ range that included the first peak was done in~\cite{Hansen:2004vq}. A comparison  of those
results with the results of our analysis is left for Section~\ref{sec:results}.

Estimating the {\it real} power spectrum from observations of
a limited region in the sky, introduces errors that increase as we 
shrink the patch. To minimize this effect we use {\it complementary masks}.
For instance, {\it to check the influence of the northern ecliptic
pole region, we just remove the cap around the pole and keep the rest of
the sky}. 

\section{Estimation of Statistical Significance}
\label{sec:stat}

%Masking a full sky map introduces error into the estimation of
%the underlying power spectrum. 
Masking a full sky map introduces correlations between the 
different multipoles that limit the precision of the estimated
underlying power spectrum.
Because our analysis of the data includes masking
large regions near the ecliptic poles, we must determine whether our results are 
statistically significant or just a result of masking the map.

To do this, we generate $10^{4}$ random realizations of the
sky map from the best fit $\Lambda CDM$ power spectrum. 
For each data release year and frequency band we wish to simulate we 
apply the corresponding beam transfer function, supplied by the WMAP team,  
to the extracted $a_{\ell m}$. 

Next, we generate two identical maps of the sky and add uncorrelated noise to the pixels of 
each, given by the function 
\begin{equation}
Noise \equiv \alpha \sigma_{\it{i}}(p),
\label{eq:noise}
\end{equation}
where 
\begin{equation}
\sigma_{\it{i}}(p)=\frac{\sigma_{0,\it{i}}}{\sqrt{N_{obs}(p)}}
\label{eq:sigma}
\end{equation}
is the noise for each pixel $p$ in the {\it{i}}th frequency band map with {\it{i}} = W or V.
 Here $\alpha$ is a randomly generated number with a Gaussian distribution,
$\sigma_{0,\it{i}}$ the mean rms noise used to calculate the pixel noise for the {\it{i}}th band provided by 
\cite{Hinshaw:2003ex}, and $N_{obs}(p)$
is the number of observations recorded for pixel $p$ that is taken from the corresponding
data release year and frequency band WMAP data map we wish to simulate. This 
mimics the noise present in the V- and W-band data maps that, being independent, 
drops from the cross-correlation
% due to the $a_{\ell m}$ cross-correlation
in Eq.~(\ref{eq:pscl}). 

Each map is then masked; first only the WMAP KQ85 mask is used, then the analysis is repeated with an 
additional region masked out (these regions include the north, south ecliptic poles and/or the ecliptic plane). We subtract the monopole and dipole,
which are induced during masking, and calculate the cross-power spectrum using Eq.~(\ref{eq:pscl}). Finally, we remove 
the window function and calculate the true $C_{\ell}s$ from the pseudo-$C_{\ell}s$ using 
Eq.~(\ref{eq:qest}).

We would  like to quantify the relative difference between a map masked with only the KQ85 mask and the 
same map masked with KQ85 + some additional region. We have thus defined a peak statistic
\begin{equation}
S_{\rm peak} \equiv 2\frac{\sum_{\ell =\ell_{min}}^{\ell_{max}}\left({C}_{\ell}^{KQ85+add'l\  mask}-{C}_{\ell}^{KQ85}\right)}
{\sum_{\ell =\ell_{min}}^{\ell_{max}}\left({C}_{\ell}^{KQ85+add'l\  mask}+{C}_{\ell}^{KQ85}\right)}.
\label{eq:peak_stat}
\end{equation}
Here $\ell_{min}$ and $\ell_{max}$ are the extrema of a range of $\ell$ that covers the first peak of the power
spectrum. The exact $\ell_{min}$=200 and $\ell_{max}$=250 values were chosen such that the value of 
$\ell_{max}(\ell_{max}+1)C_{\ell_{max}}$ and $\ell_{min}(\ell_{min}+1)C_{\ell_{min}}$ are 90\% of the WMAP five year power spectrum
peak value at its maximum.

\section{Results and discussion}
\label{sec:results}

The observed power spectrum in the first year WMAP data
for scales $\ell \sim 200$ was $\sim$10\% smaller when using only the data from the ecliptic
poles than what the best fit $\Lambda$CDM model would indicate. 
If this dip originated due to the lack of power entirely in a particular
region in the sky, we can then estimate the diameter of the circular patch involved
to be $\sim 30^\circ$. 

\begin{figure}[htp]
\begin{center}
\subfigure[First year data.]{\label{fig:powspec-1a}\includegraphics[scale=.43]{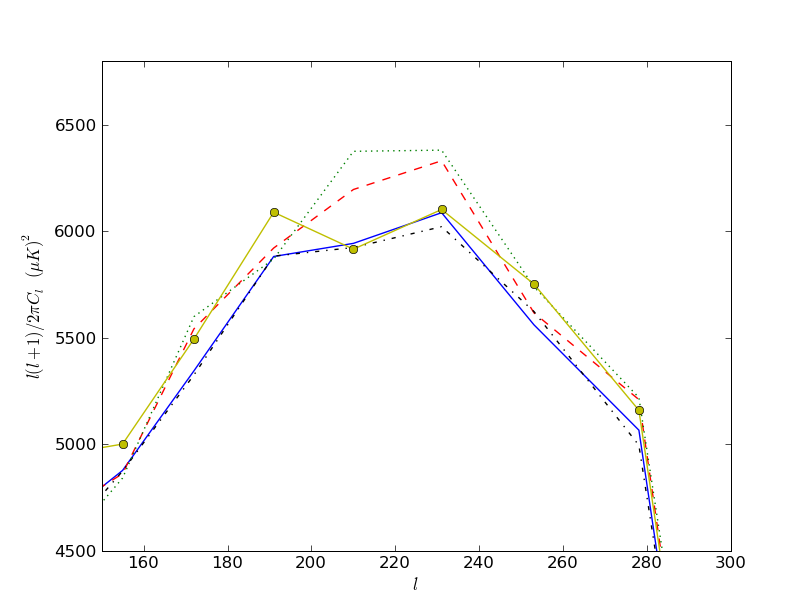}}
\subfigure[Third year data.]{\label{fig:powspec-1b}\includegraphics[scale=.43]{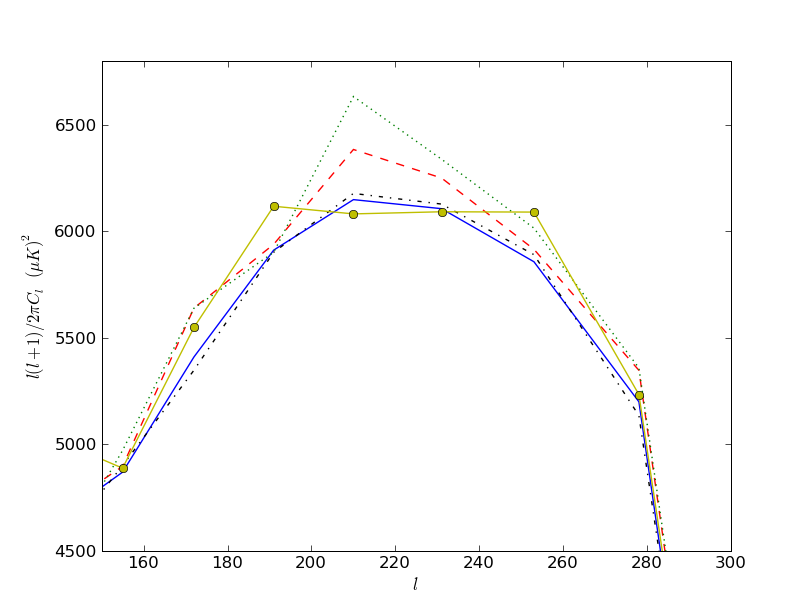}}
\subfigure[Fifth year data.]{\label{fig:powspec-1c}\includegraphics[scale=.43]{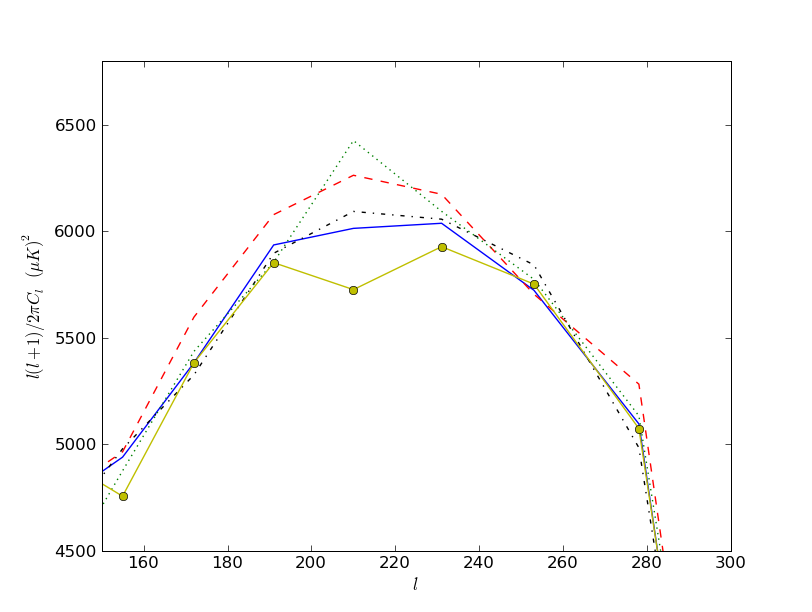}}

\end{center}
\caption{Power spectra for each data release year using the first year weighting
schemes. We show the power spectra extracted from the full sky (blue solid),
full sky without the north (red dashed) and south (black dot dashed) ecliptic caps, the ecliptic plane only (green dotted),
and the ecliptic poles only (yellow circles). The galaxy is always masked out.}
\label{fig:powspec}
\end{figure}

\begin{figure}[htp]
\begin{center}
\subfigure[First year data.]{\label{fig:powspec-3a}\includegraphics[scale=.43]{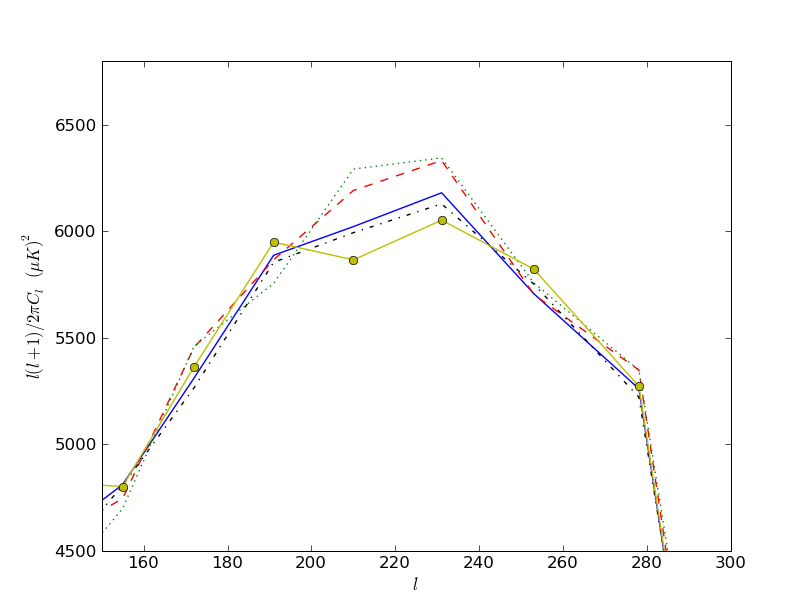}}
\subfigure[Third year data.]{\label{fig:powspec-3b}\includegraphics[scale=.43]{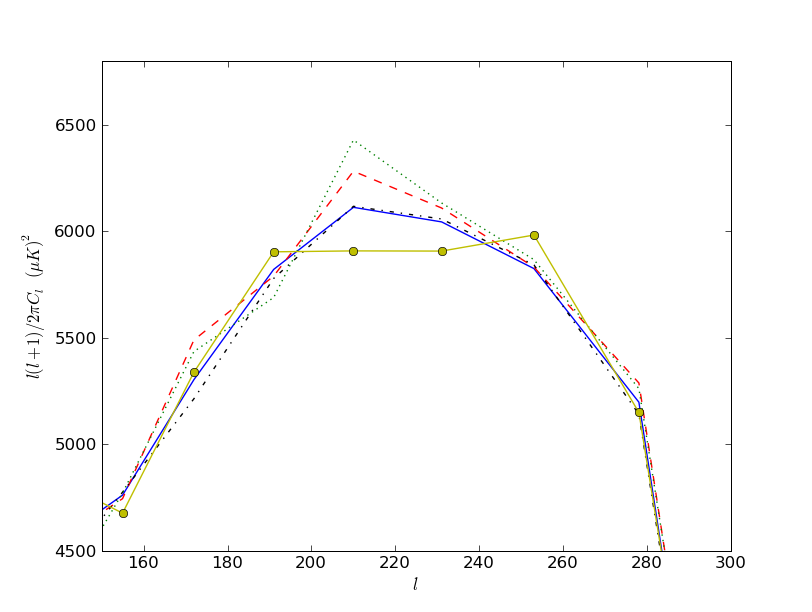}}
\subfigure[Fifth year data.]{\label{fig:powspec-3c}\includegraphics[scale=.43]{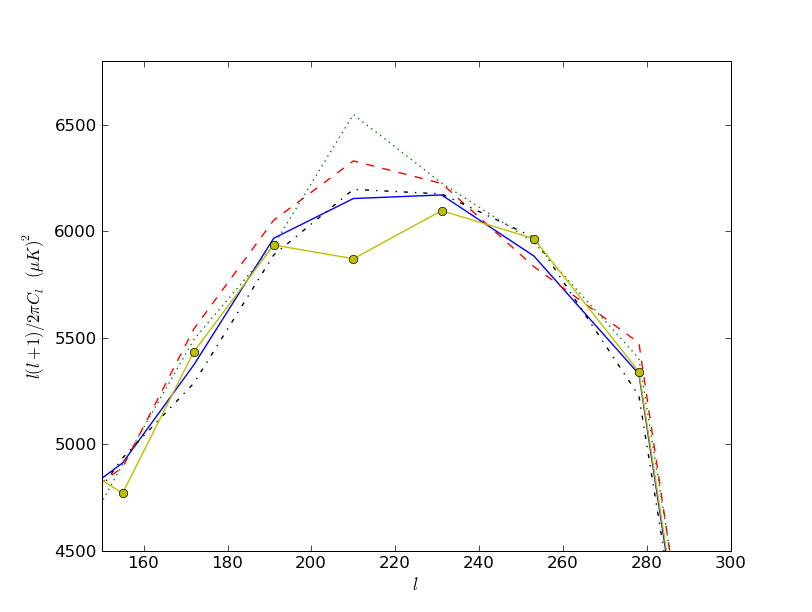}}

\end{center}
\caption{Power spectra for each data release year using the third year weighting
schemes. We show the power spectra extracted from the full sky (blue solid),
full sky without the north (red dashed) and south (black dot dashed) ecliptic caps, the ecliptic plane only (green dotted),
and the ecliptic poles only (yellow circles). The galaxy is always masked out.}
\label{fig:powspec-3}
\end{figure}

\begin{figure}[htp]
\begin{center}
\subfigure[First year weighting.]{\label{fig:powspec-1c}\includegraphics[scale=.43]{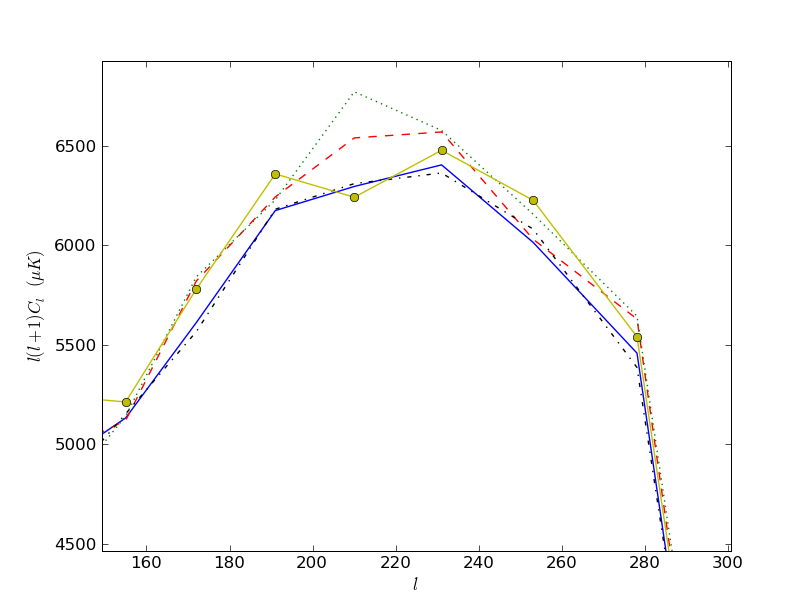}}
\subfigure[Third year weighting.]{\label{fig:powspec-3c}\includegraphics[scale=.43]{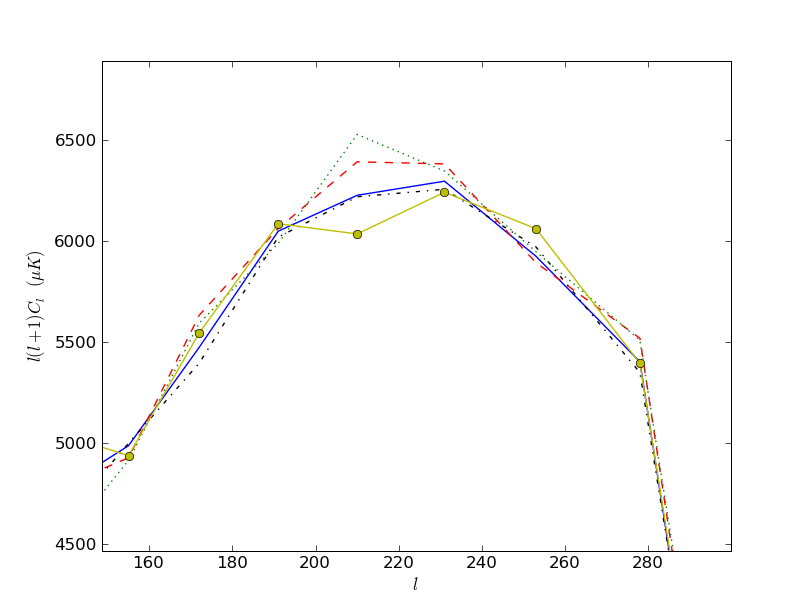}}
\end{center}
\caption{Power spectra for the full 5-year coadded data using both the first and third year weighting scheme. 
We show the power spectra extracted from the full sky (blue solid),
full sky without the north (red dashed) and south (black dot dashed) ecliptic caps, the ecliptic plane only (green dotted),
and the ecliptic poles only (yellow circles). The galaxy is always masked out.}
\label{fig:powspec-5coadd}
\end{figure}

As shown in Fig. \ref{fig:powspec}, {\it the power around} $\ell \sim 200$ 
{\it increases if data in the region of the north ecliptic pole is masked out}
when estimating the spectrum. The blue solid line was obtained with the KQ85 mask only,
and for the black dot dashed line we left out, in addition, the southern ecliptic pole. When
leaving out the northern ecliptic pole (red dashed) or keeping just the ecliptic 
plane (green dotted) the value of $C_{\ell \sim 200}$ is roughly 10\% higher.

This effect is seen {\it for all WMAP data analyzed here regardless of
the chosen weighting scheme} as shown in Fig.~\ref{fig:powspec-3} for the third year weighting scheme analysis and, 
more importantly, in Fig.~\ref{fig:powspec-5coadd} where we have used the full 5 year coadded data. Although, 
the dip-like shape in the full sky minus the galaxy only appears when the first year data and weighting scheme is used. 
The 5 year coadded data is the most compelling part of the analysis, since there are far more observations per pixel
and therefore less noise is in the resulting maps.

\begin{figure}[htp]
\begin{center}
\subfigure[]{\label{fig:peak_stat-1a}\includegraphics[scale=.43]{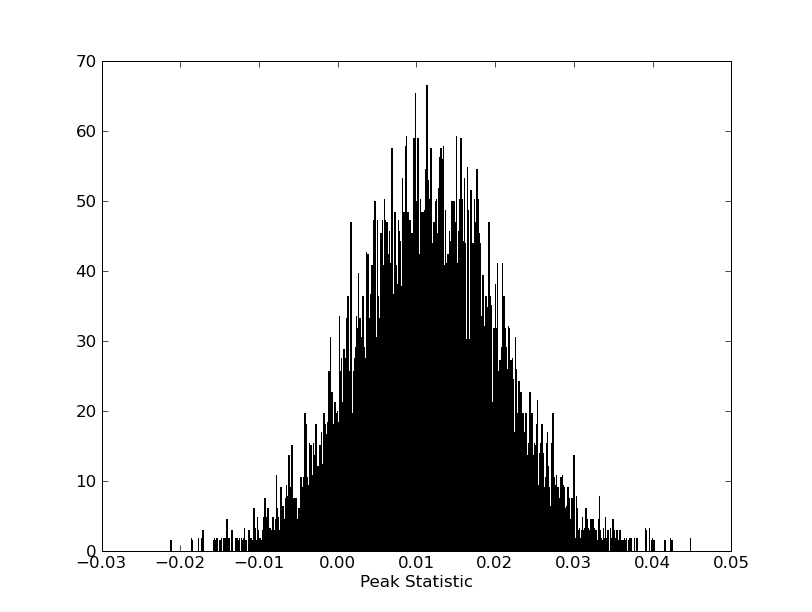}}
\subfigure[]{\label{fig:peak_stat-1b}\includegraphics[scale=.43]{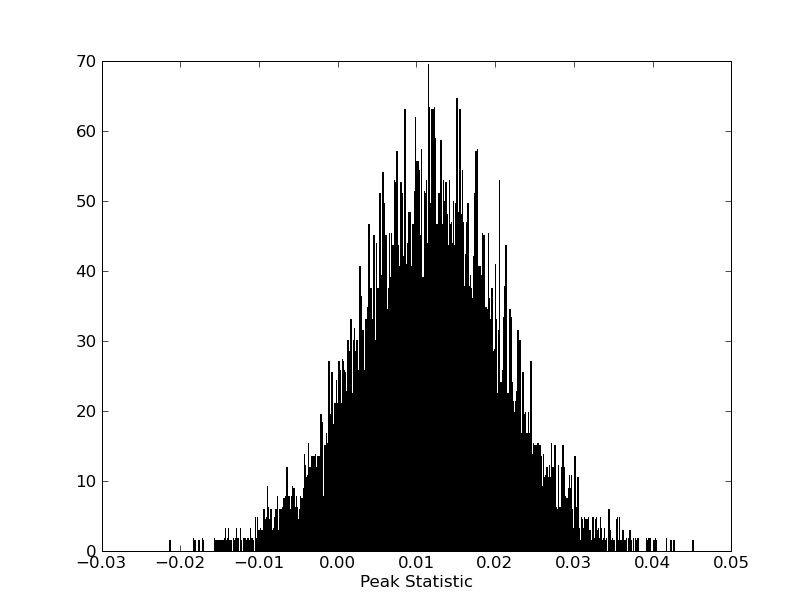}}
\subfigure[]{\label{fig:peak_stat-1c}\includegraphics[scale=.43]{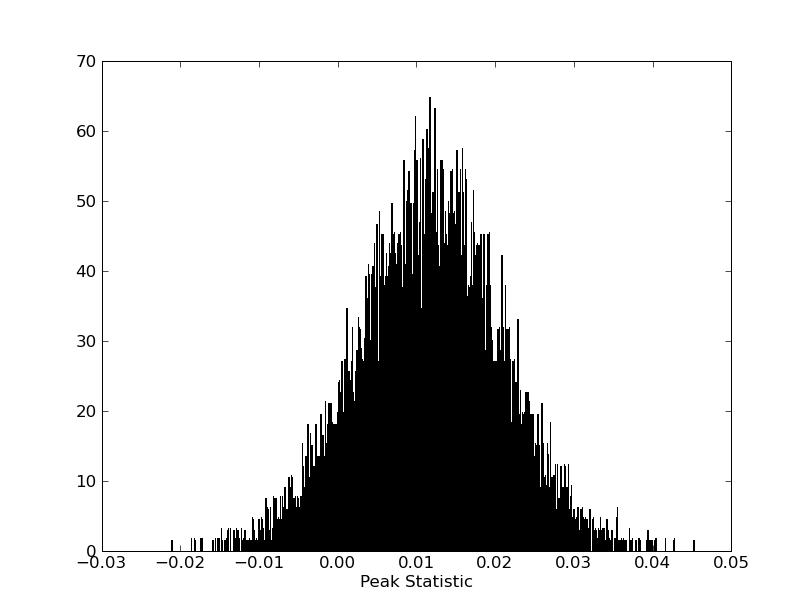}}
\end{center}
\caption{Probability density of the peak statistics calculataed using first \ref{fig:peak_stat-1a},
third \ref{fig:peak_stat-1b}, and fifth \ref{fig:peak_stat-1c} year simulated data
for the first year weighting scheme. 
The y-axis shows the normalized number of Monte Carlo skies that have each particular
 value of S.}
\label{fig:peak_stat_wscheme1yr}
\end{figure}

\begin{figure}[htp]
\begin{center}
\subfigure[]{\label{fig:peak_stat-2a}\includegraphics[scale=.43]{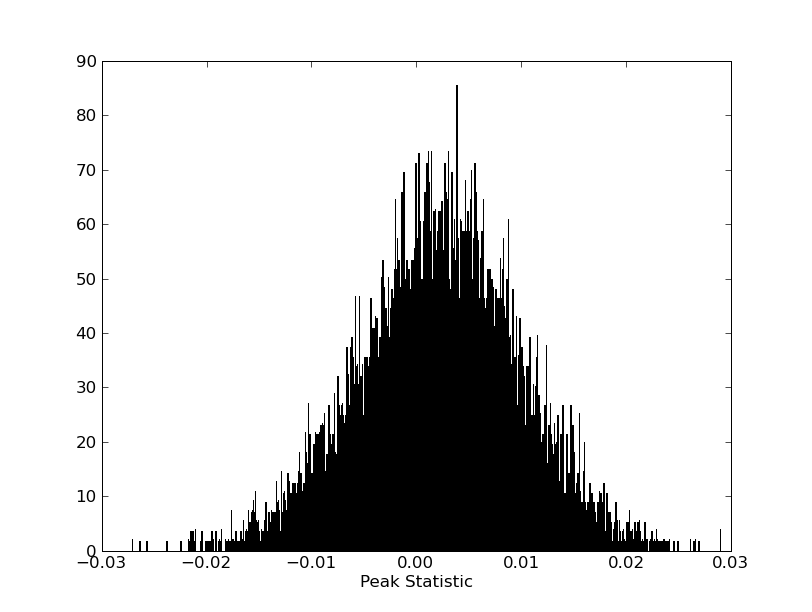}}
\subfigure[]{\label{fig:peak_stat-2b}\includegraphics[scale=.43]{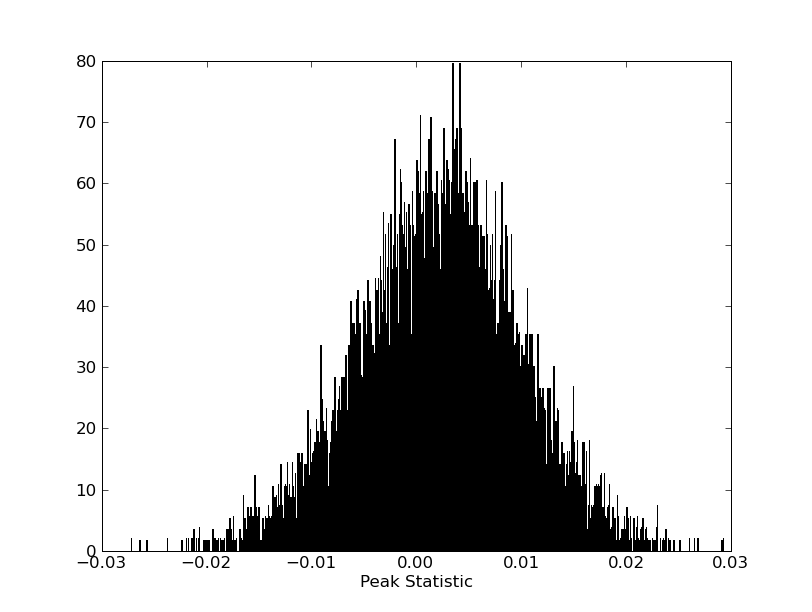}}
\subfigure[]{\label{fig:peak_stat-2c}\includegraphics[scale=.43]{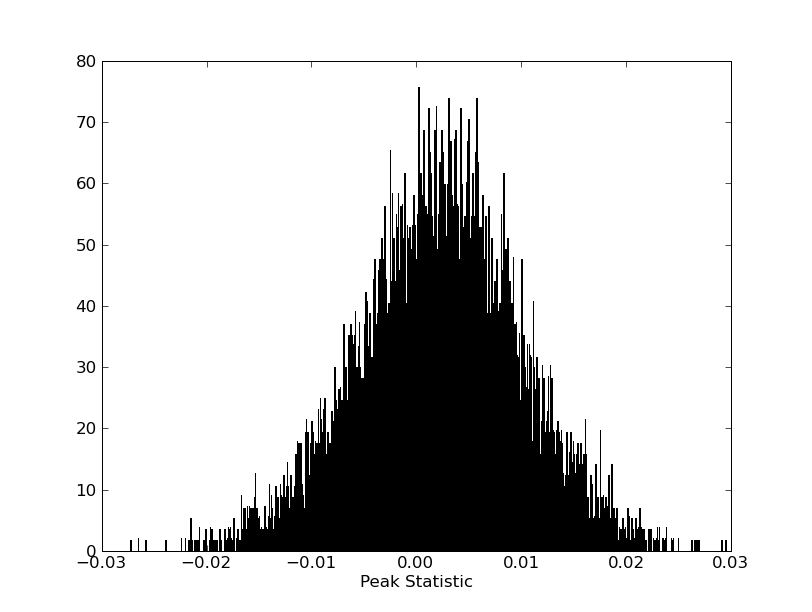}}
\end{center}
\caption{Probability density of the peak statistics calculataed using first \ref{fig:peak_stat-2a},
third \ref{fig:peak_stat-2b}, and fifth \ref{fig:peak_stat-2c} year simulated data
for the third year weighting scheme. 
The y-axis shows the normalized number of Monte Carlo skies that have each particular
 value of S.}
\label{fig:peak_stat_wscheme3yr}
\end{figure}

\begin{figure}[htp]
\begin{center}
\subfigure[First year weighting]{\label{fig:peak_stat-3a}\includegraphics[scale=.43]{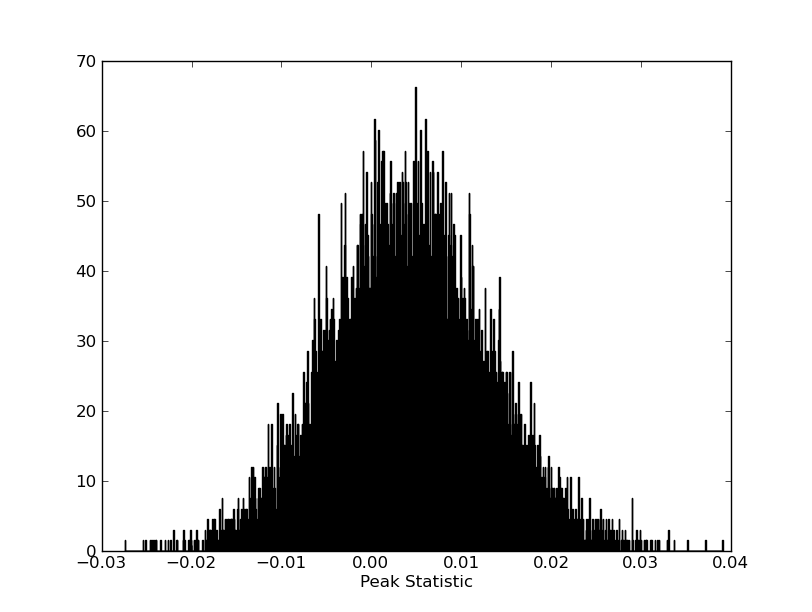}}
\subfigure[Third year weighting]{\label{fig:peak_stat-3b}\includegraphics[scale=.43]{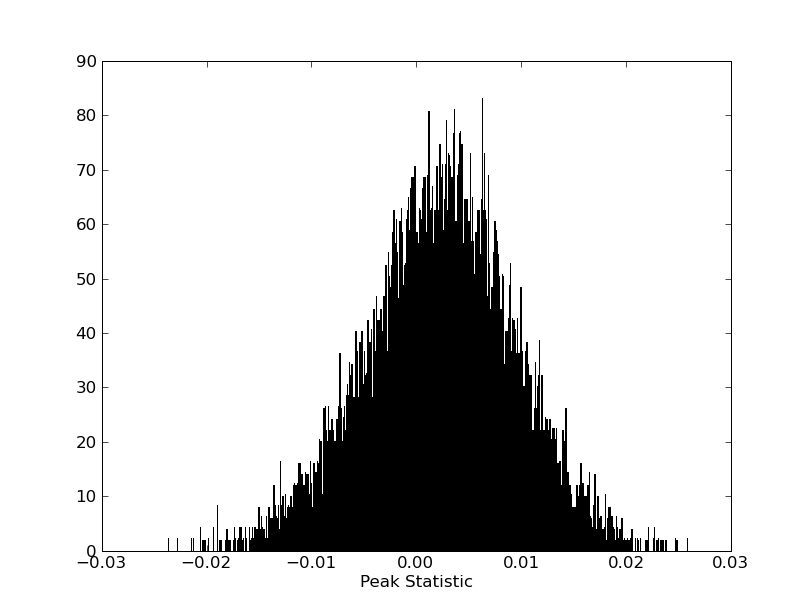}}
\end{center}
\caption{Probability density of the peak statistics calculataed using the five year coadded simulated data with the first 
\ref{fig:peak_stat-3a},
third \ref{fig:peak_stat-3b} year weighting schemes. 
The y-axis shows the normalized number of Monte Carlo skies that have each particular
 value of S.}
\label{fig:peak_stat_5yr_coadd}
\end{figure}

The significance of this effect can only be assessed with a detailed likelihood analysis. To this 
end, we have calculated the p-value for the peak statistic, given by Eq.~{\ref{eq:peak_stat}}, 
of each data release and weighting
scheme. Figs.~{\ref{fig:peak_stat_wscheme1yr}, \ref{fig:peak_stat_wscheme3yr}, \& \ref{fig:peak_stat_5yr_coadd}} 
show the probability densities of peak statistic values found for the simulations masked with the KQ85 and north 
ecliptic+KQ85 masks for the first and third year weighting schemes respectively. The peak statistics
and corresponding p-values (shown as percentages) for each data set are listed in Table~{\ref{tab:p-val}}.
These percentages are calculated by integrating under the curve in Figs. ~{\ref{fig:peak_stat_wscheme1yr}},
~{\ref{fig:peak_stat_wscheme3yr}}, \& \ref{fig:peak_stat_5yr_coadd} to the 
left of the specific value of the peak statistic found for the data.
The significance is above $\sim$95\% regardless of the data set and weighting scheme.

The process of foreground cleaning was overhauled in the third year WMAP data
release, and could account for the few percent drop in the significance value from the first to third year
analysis. Still, the reduction in power due to northern ecliptic pole data
is still significant (by about $2\sigma$) in the later data releases.

As for possible systematic
effects, it should be noted that the WMAP surveys the ecliptic plane more
sparsely than the poles, and it has been suggested that beam ellipticity
could bias the result. However, Wehus et al. \cite{Wehus:2009zh} performed
an extensive analysis of the effects of asymmetric beam patterns on the power spectrum, and found
the change to be much smaller than what is shown in Figs. \ref{fig:powspec}, \ref{fig:powspec-3} \& \ref{fig:powspec-5coadd} .

Hansen, et. al~\cite{Hansen:2004vq} performed an analysis on the first year WMAP data by varying the location of 
$9.5^{\circ}$ and $19^{\circ}$ disks around the sky and computing the power spectrum. They found that 3
disks resulted in a $2\sigma$ result, however this happened in 10\% of their simulations which lead them to 
conclude that the effect was not significant. One important difference between the Hansen, et. al analysis and our own is that
they used only the map data from inside the disk and consequently had use large $\ell$
bins to reconstruct the power spectrum. Our method, which uses complementary masks -- analysis is on the data outside
of the disc rather than inside, allows us to use a much smaller range of $\ell$ for each bin. Therefore, we are able to 
resolve the first peak in the power spectrum with a higher degree of accuracy (we have 6 points for $150<\ell<250$ whereas
they have 2). We believe this could account for the difference in the determined significance of this effect.

\begin{table}[htp]
\caption{The first columns under the weight scheme headings list the peak statistic 
calculated for each data release. The second columns list the corresponding significance.}
\centering
\begin{tabular}{|r||c|c||c|c|}
\hline
& \multicolumn{2}{|c|}{1st year weighting} & \multicolumn{2}{|c|}{3rd year weighting} \\
\hline
1-year & .0347 & 99.57\% & .0206 & 99.42\% \\ \hline
3-year & .0274 & 96.70\% & .0166 & 97.51\% \\ \hline
5-year & .0278 & 96.94\% & .0154 & 96.35\% \\ \hline
5-year coadded & .0247 & 98.89\% & .0136 & 95.61\% \\ 
\hline

\end{tabular}
\label{tab:p-val}
\end{table}
\section{Conclusion}
\label{sec:conclusion}
In the first year WMAP TT power spectrum, the binned $C_{\ell}$ value at the first peak
was significantly lower than the best-fit $\Lambda$CDM value \cite{Hinshaw:2003ex}. This was widely referred to 
as the ``dip in the first peak." In the third year data release the 
so-called dip disappeared. This was attributed to the use of an improved weighting
scheme \cite{Hinshaw:2006ia}. However, there was also shown to be a decrease in power at the first peak 
from the part of the sky near the ecliptic poles compared to the ecliptic plane in the first year 
data \cite{Hinshaw:2003ex}. Indeed, 
using the improved (third year) weighting scheme, we have found that the ecliptic polar 
power at the first peak
is significantly lower than the full-sky power in the first, third, fifth and coadded 5 year data
releases. (We have not yet analyzed the seventh year data release, although we do not expect much difference.) 

By separately masking out the north ecliptic polar region, the south ecliptic polar region 
and the ecliptic plane we have demonstrated that this difference can be ascribed to a large 
reduction in peak power in the north ecliptic polar region. We have shown that there is 
a significant increase in power when we mask out the north ecliptic pole for all 
weighting schemes and data releases. There is no such change in power when the south 
ecliptic pole is masked. Additionally, when using the third year weighting scheme 
there is still a sizeable reduction in power	when only the data from the ecliptic poles is used
compared to when the full sky is analyzed.

We have characterized the absence of power at the first doppler peak in the temperature power spectrum
in the north ecliptic polar region using the peak 
statistic defined in Eq. \ref{eq:peak_stat}. Using either the first or third year weighting
scheme and across all data releases the significance is at least 95\% based on $10^{4}$ 
best-fit $\Lambda$CDM realizations and the analysis of one $V \times W$ cross-band power 
spectrum. The most trusted result should arguably be the significance of the coadded map with the third year weight
scheme, as it provides analysis with the least noise and data processing method closest to what is currently used.
Even when analyzing the most optimal data set for characterizing this effect, our result is 
still $\approx 2\sigma$ significant.

Indeed, the first peak may seem like a specific choice of range in $\ell$ to study -- we would like
to extend this analysis in the future to other ``special" regions in $\ell$-space, namely to the second and 
third doppler peaks. However current experiments are limited in their measurements around these peaks, so
 an accurate characterization of this effect is not possible. Future CMB experiments should provide better data for
further analysis.
	
	Archeops's observation of a dip in the 
first peak from measurements including the north ecliptic polar region of the sky, 
albeit at low statistical significance, suggests that this is not a WMAP systematic. Planck
should therefore confirm the existence of a dip in the first peak power near the north 
ecliptic pole. 

If these variations are of cosmological origin, they cannot be 
explained within the standard $\Lambda$CDM model,
since it predicts an isotropic and Gaussian spectrum. However, the association with the north
ecliptic pole argues against a cosmological origin and in favor of Solar System physics. Possible
signatures include a deviation in the CMB spectrum in this region of the sky and an associated
anomaly in polarization maps.

\begin{acknowledgments}
The authors were supported by a grant from the U.S. DOE to the Particle Astrophysics theory group
at Case Western Reserve University and by NASA grant NNX07AG89G.
F.F. was supported in part by the U.S. DOE under Contract
No. DE-FG02-91ER40628 and the NSF under Grant No. PHY-0855580. 
\end{acknowledgments}


\begin{thebibliography}{}
%\cite{Hinshaw:2003ex}
\bibitem{Hinshaw:2003ex}
  G.~Hinshaw {\it et al.}  [WMAP Collaboration],
  %``First Year Wilkinson Microwave Anisotropy Probe (WMAP) Observations:
  %Angular Power Spectrum,''
  Astrophys.\ J.\ Suppl.\  {\bf 148}, 135 (2003).
%  [arXiv:astro-ph/0302217].
  %%CITATION = APJSA,148,135;%%

%\cite{Spergel:2003cb}
\bibitem{Spergel:2003cb}
  D.~N.~Spergel {\it et al.}  [WMAP Collaboration],
  %``First Year Wilkinson Microwave Anisotropy Probe (WMAP) Observations:
  %Determination of Cosmological Parameters,''
  Astrophys.\ J.\ Suppl.\  {\bf 148}, 175 (2003).
%  [arXiv:astro-ph/0302209].
  %%CITATION = APJSA,148,175;%%

%\cite{Hinshaw:2006ia}
\bibitem{Hinshaw:2006ia}
  G.~Hinshaw {\it et al.}  [WMAP Collaboration],
  %``Three-year Wilkinson Microwave Anisotropy Probe (WMAP) observations:
  %Temperature analysis,''
  Astrophys.\ J.\ Suppl.\  {\bf 170}, 288 (2007).
%  [arXiv:astro-ph/0603451].
  %%CITATION = APJSA,170,288;%%
  
%\cite{Bennett:2010jb}
\bibitem{Bennett:2010jb}
  C.~L.~Bennett {\it et al.},
  %``Seven-Year Wilkinson Microwave Anisotropy Probe (WMAP) Observations: Are
  %There Cosmic Microwave Background Anomalies?,''
  arXiv:1001.4758 [astro-ph.CO].
  %%CITATION = ARXIV:1001.4758;%%

%\cite{Copi:2010na}
\bibitem{Copi:2010na}
  C.~J.~Copi, D.~Huterer, D.~J.~Schwarz and G.~D.~Starkman,
  %``Large angle anomalies in the CMB,''
  arXiv:1004.5602 [astro-ph.CO].
  %%CITATION = ARXIV:1004.5602;%%


%\cite{de OliveiraCosta:2003pu}
\bibitem{de OliveiraCosta:2003pu}
  A.~de Oliveira-Costa, M.~Tegmark, M.~Zaldarriaga and A.~Hamilton,
  %``The significance of the largest scale CMB fluctuations in WMAP,''
  Phys.\ Rev.\  D {\bf 69}, 063516 (2004).
%  [arXiv:astro-ph/0307282].
  %%CITATION = PHRVA,D69,063516;%%


%\cite{Schwarz:2004gk}
\bibitem{Schwarz:2004gk}
  D.~J.~Schwarz, G.~D.~Starkman, D.~Huterer and C.~J.~Copi,
  %``Is the low-l microwave background cosmic?,''
  Phys.\ Rev.\ Lett.\  {\bf 93}, 221301 (2004).
%  [arXiv:astro-ph/0403353].
  %%CITATION = PRLTA,93,221301;%%
  
%\cite{Land:2005ad}
\bibitem{Land:2005ad}
  K.~Land and J.~Magueijo,
  %``The axis of evil,''
  Phys.\ Rev.\ Lett.\  {\bf 95}, 071301 (2005)
  [arXiv:astro-ph/0502237].
  %%CITATION = PRLTA,95,071301;%%


%\cite{Eriksen:2003db}
\bibitem{Eriksen:2003db}
  H.~K.~Eriksen, F.~K.~Hansen, A.~J.~Banday, K.~M.~Gorski and P.~B.~Lilje,
  %``Asymmetries in the CMB anisotropy field,''
  Astrophys.\ J.\  {\bf 605}, 14 (2004)
  [Erratum-ibid.\  {\bf 609}, 1198 (2004)].
%  [arXiv:astro-ph/0307507].
  %%CITATION = ASJOA,605,14;%%
  

%\cite{Rath:2007ti}
\bibitem{Rath:2007ti}
  C.~R{\"a}th, P.~Schuecker and A.~J.~Banday,
  %``Model-Independent Test for Scale-Dependent Non-Gaussianities in the Cosmic
  %Microwave Background,''
  Phys.\ Rev.\ Lett.\  {\bf 102}, 131301 (2009)
  [arXiv:astro-ph/0702163].
  %%CITATION = PRLTA,102,131301;%%

%\cite{Rossmanith:2009cy}
\bibitem{Rossmanith:2009cy}
  G.~Rossmanith, C.~R{\"a}th, A.~J.~Banday and G.~Morfill,
  %``Non-Gaussian Signatures in the five-year WMAP data as identified with
  %isotropic scaling indices,''
  arXiv:0905.2854 [astro-ph.CO].
  %%CITATION = ARXIV:0905.2854;%%

  
  %\cite{Hansen:2008ym}
\bibitem{Hansen:2008ym}
  F.~K.~Hansen, A.~J.~Banday, K.~M.~Gorski, H.~K.~Eriksen and P.~B.~Lilje,
  %``Power Asymmetry in Cosmic Microwave Background Fluctuations from Full Sky
  %to Sub-degree Scales: Is the Universe Isotropic?,''
  Astrophys.\ J.\  {\bf 704}, 1448 (2009)
  [arXiv:0812.3795 [astro-ph]].
  %%CITATION = ASJOA,704,1448;%%
  
    %\cite{Hansen:2004vq}
\bibitem{Hansen:2004vq}
  F.~K.~Hansen, A.~J.~Banday and K.~M.~Gorski,
  %``Testing the cosmological principle of isotropy: local power spectrum
  %estimates of the WMAP data,''
  Mon.\ Not.\ Roy.\ Astron.\ Soc.\  {\bf 354}, 641 (2004)
  [arXiv:astro-ph/0404206].
  %%CITATION = MNRAA,354,641;%%

%\cite{Vielva:2003et}
\bibitem{Vielva:2003et}
  P.~Vielva, E.~Martinez-Gonzalez, R.~B.~Barreiro, J.~L.~Sanz and L.~Cayon,
  %``Detection of non-Gaussianity in the WMAP 1-year data using spherical
  %wavelets,''
  Astrophys.\ J.\  {\bf 609}, 22 (2004)
  [arXiv:astro-ph/0310273].
  %%CITATION = ASJOA,609,22;%%


%\cite{Cruz:2004ce}
\bibitem{Cruz:2004ce}
  M.~Cruz, E.~Martinez-Gonzalez, P.~Vielva and L.~Cayon,
  %``Detection of a non-Gaussian Spot in WMAP,''
  Mon.\ Not.\ Roy.\ Astron.\ Soc.\  {\bf 356}, 29 (2005)
  [arXiv:astro-ph/0405341].
  %%CITATION = MNRAA,356,29;%%

%\cite{Yadav:2007yy}
\bibitem{Yadav:2007yy}
  A.~P.~S.~Yadav and B.~D.~Wandelt,
  %``Evidence of Primordial Non-Gaussianity $(f_{\rm NL})$ in the Wilkinson
  %Microwave Anisotropy Probe 3-Year Data at 2.8$\sigma$,''
  Phys.\ Rev.\ Lett.\  {\bf 100}, 181301 (2008)
  [arXiv:0712.1148 [astro-ph]].
  %%CITATION = PRLTA,100,181301;%%


%\cite{Komatsu:2008hk}
\bibitem{Komatsu:2008hk}
  E.~Komatsu {\it et al.}  [WMAP Collaboration],
  %``Five-Year Wilkinson Microwave Anisotropy Probe (WMAP\altaffilmark 1 )
  %Observations:Cosmological Interpretation,''
  Astrophys.\ J.\ Suppl.\  {\bf 180}, 330 (2009)
  [arXiv:0803.0547 [astro-ph]].
  %%CITATION = APJSA,180,330;%%

%\cite{Smith:2009jr}
\bibitem{Smith:2009jr}
  K.~M.~Smith, L.~Senatore and M.~Zaldarriaga,
  %``Optimal limits on f_{NL}^{local} from WMAP 5-year data,''
  JCAP {\bf 0909}, 006 (2009)
  [arXiv:0901.2572 [astro-ph]].
  %%CITATION = JCAPA,0909,006;%%

%\cite{Copi:2006tu}
\bibitem{Copi:2006tu}
  C.~Copi, D.~Huterer, D.~Schwarz and G.~Starkman,
  %``The Uncorrelated Universe: Statistical Anisotropy and the Vanishing Angular
  %Correlation Function in WMAP Years 1-3,''
  Phys.\ Rev.\  D {\bf 75}, 023507 (2007).
%  [arXiv:astro-ph/0605135].
  %%CITATION = PHRVA,D75,023507;%%


%\cite{Eriksen:2007pc}
\bibitem{Eriksen:2007pc}
  H.~K.~Eriksen, A.~J.~Banday, K.~M.~Gorski, F.~K.~Hansen and P.~B.~Lilje,
  %``Hemispherical power asymmetry in the three-year Wilkinson Microwave
  %Anisotropy Probe sky maps,''
  Astrophys.\ J.\  {\bf 660}, L81 (2007).
%  [arXiv:astro-ph/0701089].
  %%CITATION = ASJOA,660,L81;%%


%\cite{Bartolo:2004if}
\bibitem{Bartolo:2004if}
  N.~Bartolo, E.~Komatsu, S.~Matarrese and A.~Riotto,
  %``Non-Gaussianity from inflation: Theory and observations,''
  Phys.\ Rept.\  {\bf 402}, 103 (2004)
  [arXiv:astro-ph/0406398].
  %%CITATION = PRPLC,402,103;%%

%\cite{Gordon:2003fi}
\bibitem{Gordon:2003fi}
  C.~Gordon and A.~Lewis,
  %``Curvaton model constraints from WMAP,''
  New Astron.\ Rev.\  {\bf 47}, 793 (2003).
  %%CITATION = ASTRE,47,793;%%

%\cite{Ferrer:2004nv}
\bibitem{Ferrer:2004nv}
  F.~Ferrer, S.~Rasanen and J.~Valiviita,
  %``Correlated isocurvature perturbations from mixed inflaton-curvaton
  %decay,''
  JCAP {\bf 0410}, 010 (2004)
  [arXiv:astro-ph/0407300].
  %%CITATION = JCAPA,0410,010;%%

%\cite{Cruz:2007pe}
\bibitem{Cruz:2007pe}
  M.~Cruz, N.~Turok, P.~Vielva, E.~Martinez-Gonzalez and M.~Hobson,
  %``A Cosmic Microwave Background feature consistent with a cosmic texture,''
  Nature {\bf 318}, 1612 (2007)
  [arXiv:0710.5737 [astro-ph]].
  %%CITATION = NATUA,318,1612;%%

%\cite{Wang:2002rta}
\bibitem{Wang:2002rta}
  X.~Wang, M.~Tegmark, B.~Jain and M.~Zaldarriaga,
  %``The last stand before MAP: cosmological parameters from lensing, CMB and
  %galaxy clustering,''
  Phys.\ Rev.\  D {\bf 68}, 123001 (2003)
  [arXiv:astro-ph/0212417].
  %%CITATION = PHRVA,D68,123001;%%

%\cite{Efstathiou:2003dj}
\bibitem{Efstathiou:2003dj}
  G.~Efstathiou,
  %``Myths and Truths Concerning Estimation of Power Spectra,''
  Mon.\ Not.\ Roy.\ Astron.\ Soc.\  {\bf 349}, 603 (2004)
  [arXiv:astro-ph/0307515].
  %%CITATION = MNRAA,349,603;%%

%\cite{Hivon:2001jp}
\bibitem{Hivon:2001jp}
  E.~Hivon, K.~M.~Gorski, C.~B.~Netterfield, B.~P.~Crill, S.~Prunet and F.~Hansen,
  %``MASTER of the CMB Anisotropy Power Spectrum: A Fast Method for Statistical
  %Analysis of Large and Complex CMB Data Sets,''
  Astrophys.\ J.\  {\bf 567}, 2 (2002)
  [arXiv:astro-ph/0105302].
  %%CITATION = ASTRO-PH/0105302;%%

%\cite{Verde:2003ey}
\bibitem{Verde:2003ey}
  L.~Verde {\it et al.}  [WMAP Collaboration],
  %``First Year Wilkinson Microwave Anisotropy Probe (WMAP) Observations:
  %Parameter Estimation Methodology,''
  Astrophys.\ J.\ Suppl.\  {\bf 148}, 195 (2003)
  [arXiv:astro-ph/0302218].
  %%CITATION = APJSA,148,195;%%
  
  %\cite{Wehus:2009zh}
\bibitem{Wehus:2009zh}
  I.~K.~Wehus, L.~Ackerman, H.~K.~Eriksen and N.~E.~Groeneboom,
  %``The effect of asymmetric beams in the Wilkinson Microwave Anisotropy Probe
  %experiment,''
  Astrophys.\ J.\  {\bf 707}, 343 (2009)
  [arXiv:0904.3998 [astro-ph.CO]].
  %%CITATION = ASJOA,707,343;%%


\end{thebibliography}
\end{document}